%% file: creynolds.tex
\documentclass{evn2004}
\usepackage{txfonts}
\usepackage{graphicx}
\usepackage{natbib}
\begin{document}
\title{Intermediate Scale Structures in BL Lac objects } 

\author{Cormac Reynolds\inst{1}
          \and
          Timothy V. Cawthorne\inst{2}
          \and
          Denise C. Gabuzda\inst{3}
}

\institute{Joint Institute for VLBI in Europe, Postbus 2, 7990 AA Dwingeloo,
        The Netherlands
         \and
        Department of Physics, Astronomy and Mathematics, University of Central
        Lancashire, Preston, Lancashire, PR1 2HE 
        \and
        Physics Department, University College Cork, Cork, Ireland
}

\abstract{

The parsec-scale total intensity structures of BL Lac objects share many
characteristics with quasars: e.g. well-collimated jets and superluminal
motion.  However, the jets of BL Lac objects appear to fade much more quickly
than those in quasars and on VLA scales many BL Lacs have structures comparable
to those of low luminosity (FR I) radio galaxies, in which kpc-scale flow
speeds are non-relativistic.

The region between parsec and kiloparsec scales is therefore one of transition
in BL Lac objects. We have carried out VLBI observations of a small sample of
BL Lac objects at frequencies between 2.3 GHz and 327 MHz in order to
investigate this transition region. Preliminary results for two of these
sources are presented here.

Coherent structures are detected in these sources out to a distance of several
tens of parsecs. We find evidence for significant changes in jet structure on
scales of a few tens of milliarcseconds, both in terms of the orientation and
collimation of the jets.

}

\maketitle
%

\section{Introduction}
BL~Lac objects are a sub-class of extragalactic radio sources exhibiting highly
variable polarized radio emission and are distinguished from optically
violently variable quasars by their weak optical line emission (equivalent
width less than 5 \AA, \citealp{stickel93}).

The parsec (pc) and kpc scale jets of BL~Lac objects are often misaligned and
the kpc scale emission is typically more diffuse. This indicates that the jet
has become decollimated between the milliarcsecond (mas) and arcsec scales.
There is evidence in some BL~Lac objects for significant total intensity and
polarized flux density on scales intermediate to the mas and arcsec resolution
provided by VLBI and VLA observations respectively \citep{gabuzda94}. Further,
the pc scale jets of BL~Lac objects exhibit inferred magnetic fields which tend
to be perpendicular to the direction of the jet motion (e.g.
\citealp{gabuzda99} and references therein), while on arcsecond scales the
magnetic field is often aligned with the jet \citep{kollgaard92}. The region
over which these changes take place lies on scales between those accessible to
high-frequency ($> 5$~GHz) VLBI and more conventional interferometers such as
the VLA, and has therefore been little explored to date. 

In order to investigate these changes of structure with scale, we have carried
out observations at frequencies of 2.3~GHz, 1.7~GHz, 610~MHz and 327~MHz with
the VLBA. At the lower frequencies more diffuse jet emission should become more
dominant allowing us to trace the development of the jet on larger scales.

Linear sizes and luminosities have been calculated using $H_0 = 75$ km s$^{-1}$ Mpc$^{-1}$, $q_0=0.05$.

\section{Observations}
The observations took place on 13th December 1999 using all ten elements of the
VLBA in a dual polarization mode at a recording rate of 128 Mb s$^{-1}$. The
data were correlated at the VLBA correlator in Socorro, New Mexico. The
observing frequencies were 2.3~GHz, 1.7~GHz, 610~MHz and 327~MHz. The 2.3~GHz
and 1.7~GHz data were processed in full polarization mode, but the polarization
maps are not presented here. The a priori calibration and self-calibration of
the data were performed using the standard programmes in the NRAO Astronomical
Image Processing Software (\texttt{AIPS}) package.

In total, a small sample of 5 objects were observed, maps for two of which are
presented here.

\section{Results}
The observed sources were inferred to have significant flux density on
intermediate scales from a comparison of their flux densities in previous VLA
and VLBI observations at a frequency of 5~GHz. The expected 5~GHz intermediate
scale flux densities of the two sources are presented in
Table~\ref{table:isflux}. For both of these sources we have been able to image
significant emission on larger scales than has been reported in previous
observations at higher frequencies.

\begin{table}
\caption[]{Intermediate scale (I.S.) flux densities inferred from
previous 5~GHz VLA and VLBI images \citep{gabuzda94,gabuzda92}.}

\label{table:isflux}
$$
\begin{array}{p{0.5\linewidth}p{0.4\linewidth}}
\hline
\noalign{\smallskip}
Source      &  5~GHz I.S. flux density (Jy) \\
\noalign{\smallskip}
\hline
\noalign{\smallskip}
1219+285  &   0.22    \\
0735+178  &   0.87    \\
\noalign{\smallskip}
\hline
\end{array}
$$
\end{table}

\subsection{1219+285}
1219+285 lies in the centre of an elliptical galaxy at a redshift of 0.102
\citep{weistrop85}. This source is essentially unresolved to the VLA D array
\citep{kollgaard92}. The VLBI jet extends to the east of the core and a number
of superluminal components have been identified \citep{gabuzda94}. 

\begin{figure}
\centering
\includegraphics[width=6.5cm]{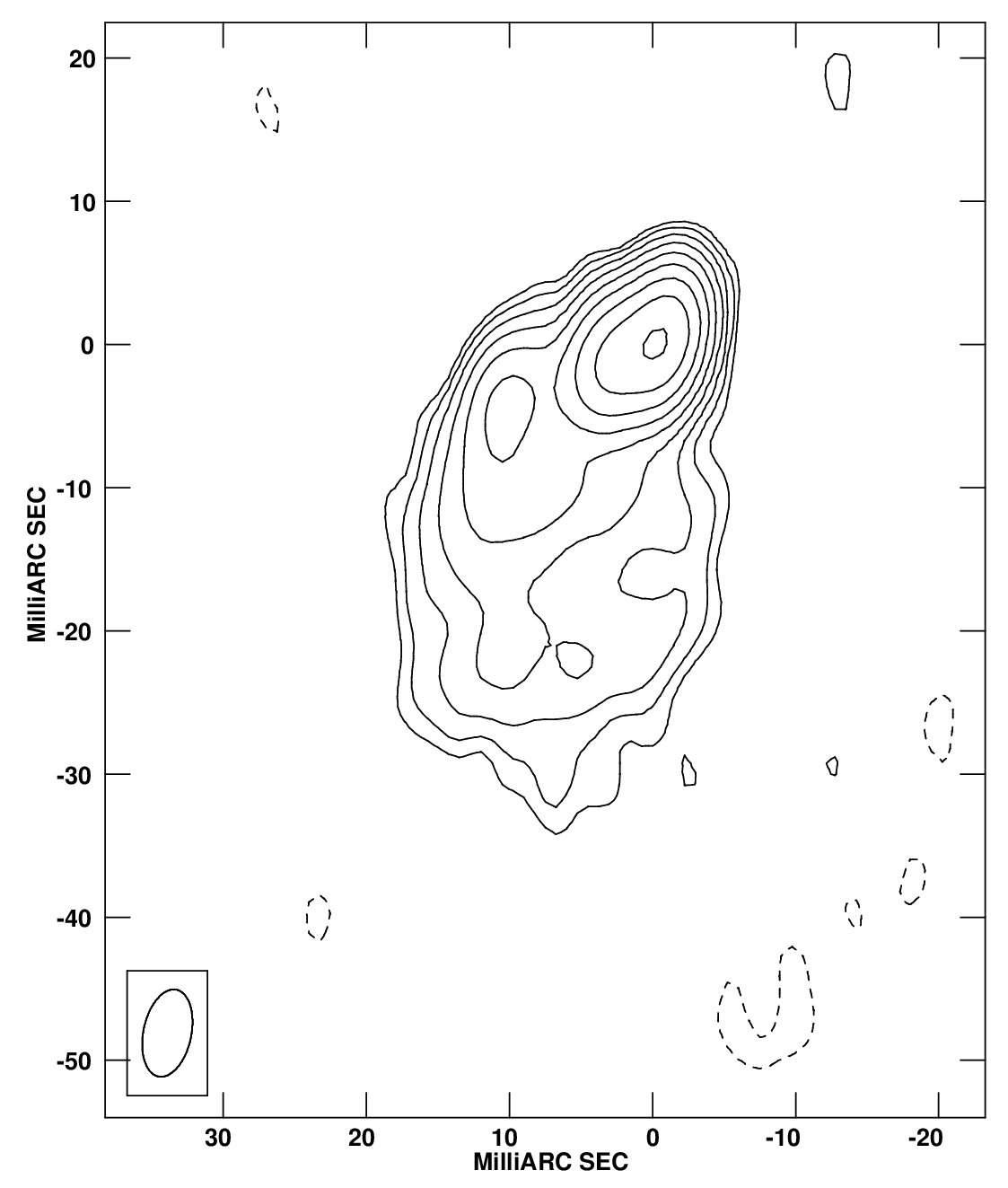}
   \caption{2.3~GHz image of 1219+285. The peak flux density is 0.38 Jy/beam
and the contour levels are $-0.36$, 0.36, 0.72, 1.44 ... 92.2 \%.
           }
      \label{fig:1219+285_2.3ghz}
\end{figure}

\begin{figure}
\centering
\includegraphics[width=6.5cm]{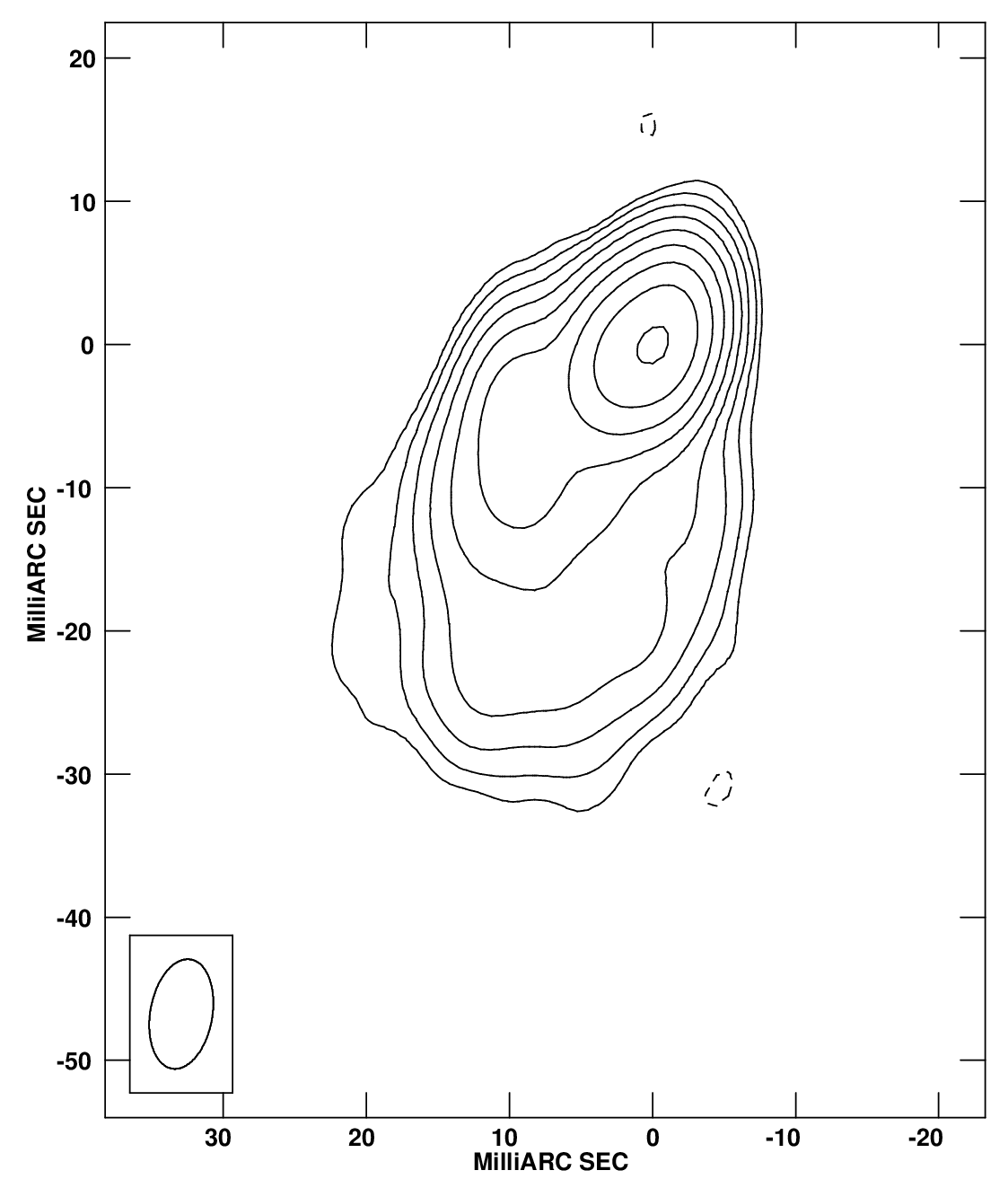}
   \caption{1.7~GHz image of 1219+285. The peak flux density is 0.38 Jy/beam
and the contour levels are $-0.36$, 0.36, 0.72, 1.44 ... 92.2 \%.
           }
      \label{fig:1219+285_1.7ghz}
\end{figure}

\begin{figure}
\centering
\includegraphics[width=6.5cm]{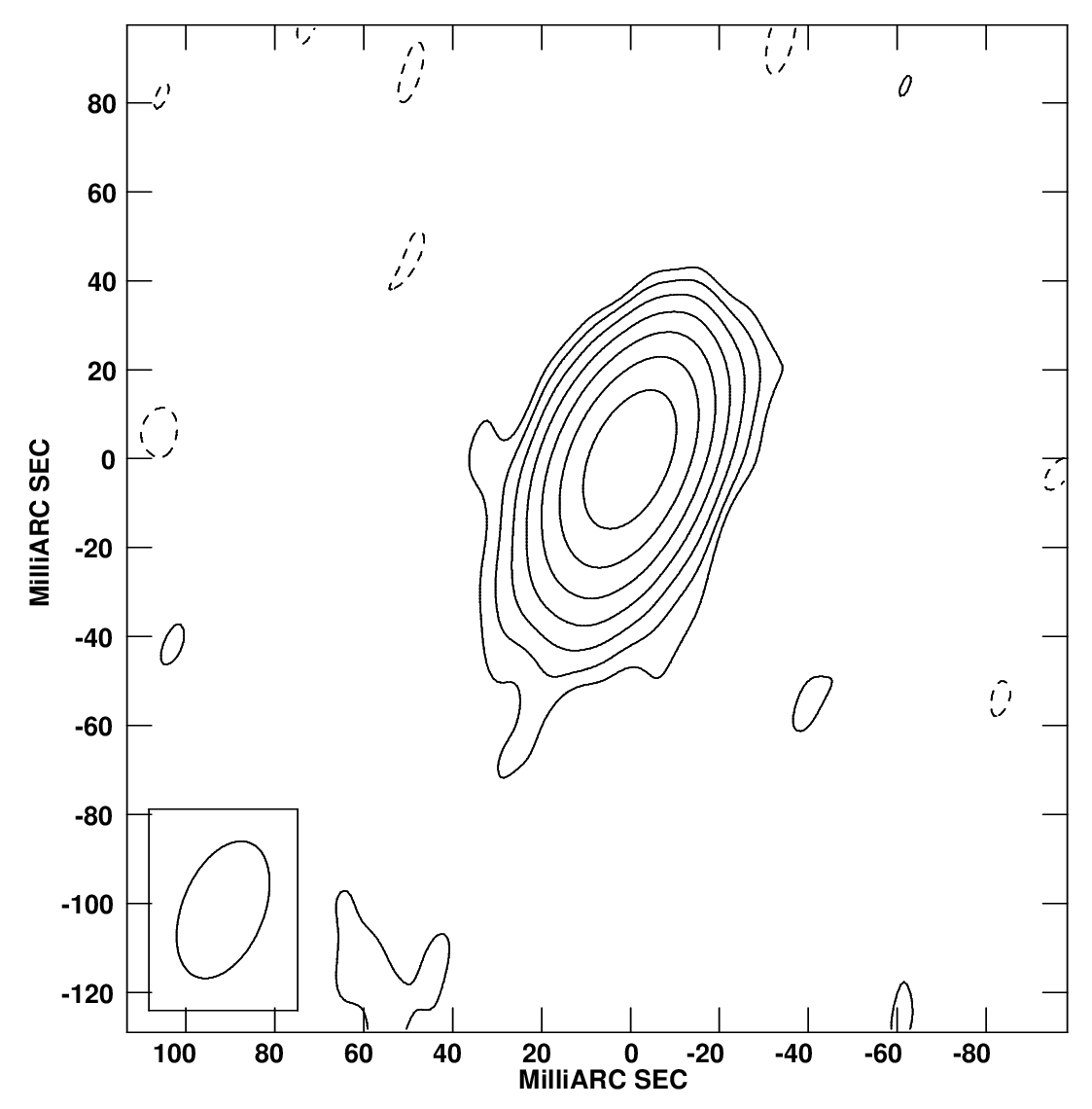}
   \caption{610~MHz image of 1219+285. The peak flux density is 0.61 Jy/beam
and the contour levels are $-0.8$, 0.8, 1.6, 3.2 ... 51.2 \%.
           }
      \label{fig:1219+285_0.6ghz}
\end{figure}

\begin{figure}
\centering
\includegraphics[width=6.5cm]{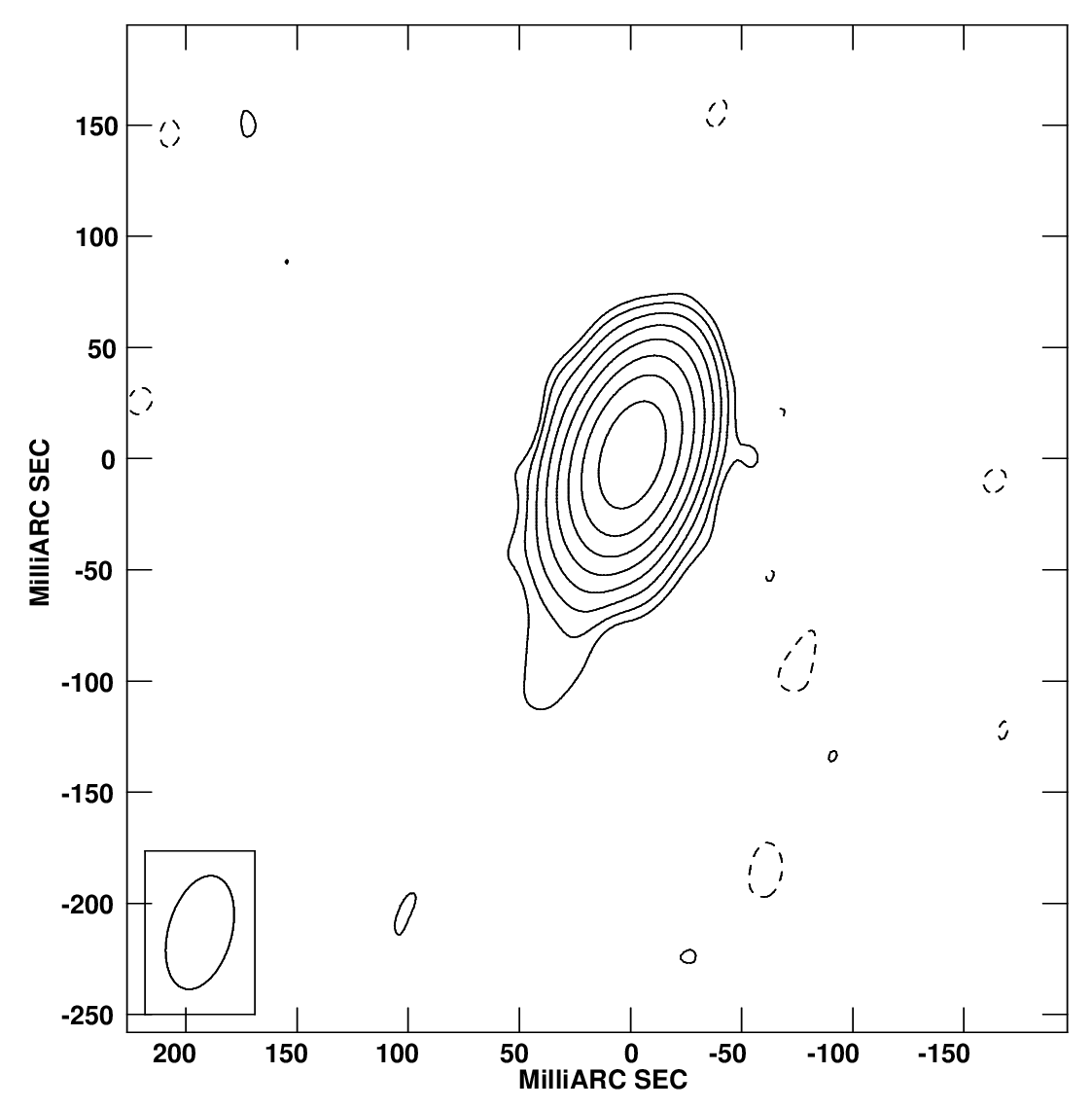}
   \caption{327~MHz image of 1219+285. The peak flux density is 0.66 Jy/beam
and the contour levels are $-0.45$, 0.45, 0.9, 1.8 ... 57.6 \%.
           }
      \label{fig:1219+285_0.3ghz}
\end{figure}

\begin{figure}
\centering
\includegraphics[width=6.5cm]{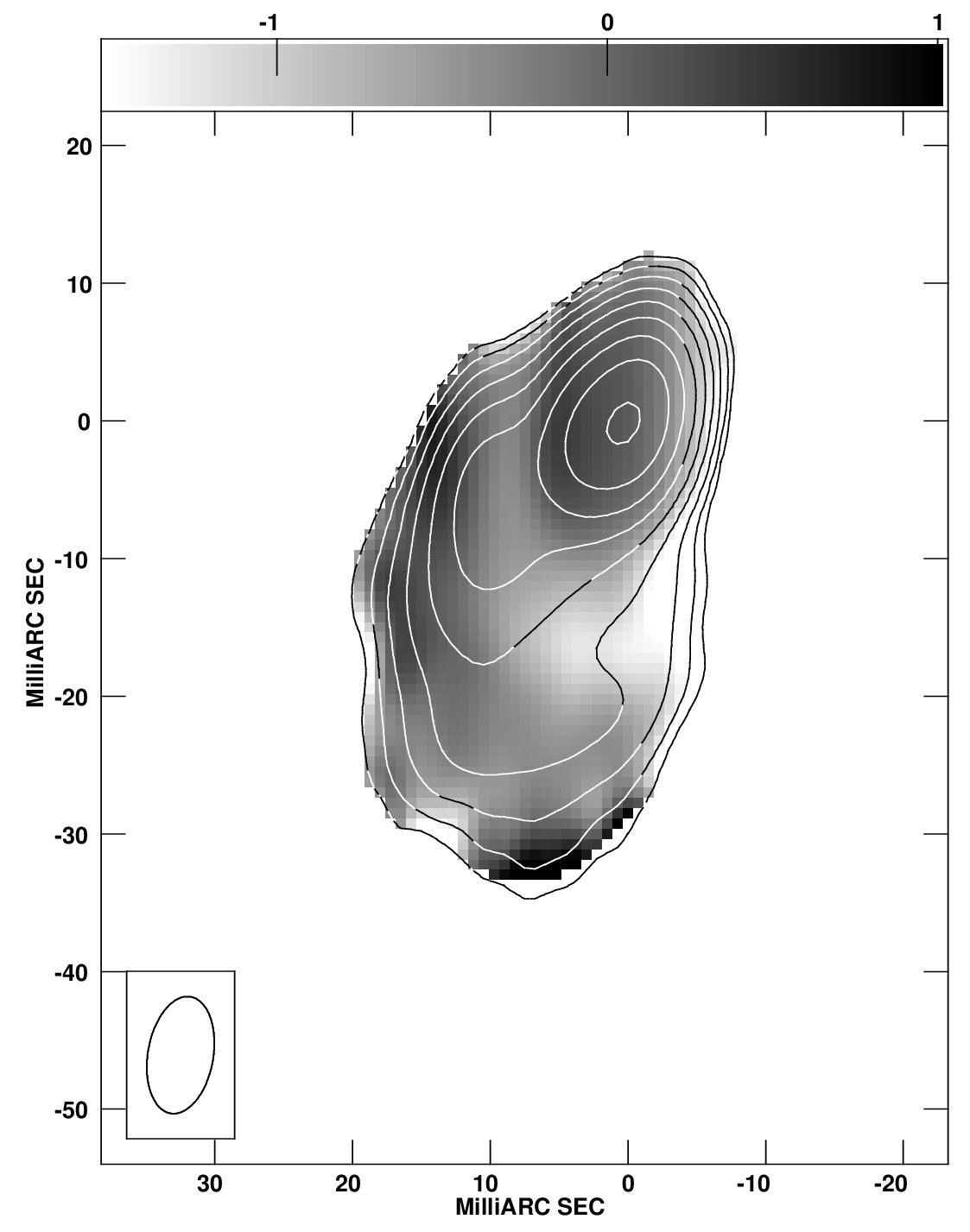}
   \caption{The contours are of the 2.3~GHz image shown in
Figure~\ref{fig:1219+285_1.7ghz}. The grey scale is the spectral index
measured between 2.3 and 1.7~GHz.
           }
      \label{fig:1219+285_specind}
\end{figure}

The images in Figures~\ref{fig:1219+285_2.3ghz} to \ref{fig:1219+285_0.3ghz}
show a large diffuse extension towards the south roughly perpendicular to the
known VLBI jet. There is an apparent change in jet direction visible in the
2.3~GHz data. It is possible that the jet is changing direction from an
easterly to a southerly direction and finally in a westerly direction, while
also becoming more diffuse. If the jet is curving towards the  line of sight
then it is also possible that this more diffuse emission from the southern part
of the source represents emission from the end of a highly projected jet.
However, it is also possible that there is a qualitative change in the jet
structure and that the jet itself is becoming diffuse and decollimated. It is
also of note that there is a region of enhanced brightness and relatively flat
spectral index (perhaps resulting from free-free absorption of a steep
spetctrum jet, see Figure~\ref{fig:1219+285_specind}) in approximately the
position that the jet direction apparently begins to change from an easterly to
a southerly direction. This may be indicative of an interaction between the jet
and the surrounding medium which results in the jet changing direction. Further
analysis of the associated polarization data should provide more information on
this point.

The expected intermediate scale flux density from comparison of VLA and VLBI
images is 0.22~Jy at 5~GHz. The 2.3~GHz image presented here detects
approximately 0.11~Jy outside the region which was observed in the previous
5~GHz VLBI observations \citep{gabuzda94}. Even allowing for spectral index
effects, this represents a large fraction of the missing 5~GHz flux density,
although it also suggests that there is significant emission on still larger
scales than these observations are sensitive to. It is interesting that no
significant new emission becomes visible in the lower frequency 610 and
327~MHz images when compared with the 1.7 and 2.3~GHz images. The sources
appear essentially unresolved at these lower frequencies.

At the redshift of this object, 1 mas $ \sim 1.71$ pc. The luminosity detected
in the new, extended structure in this observation is $\sim 2.2 \times 10^{24}$
W Hz$^{-1}$.

\subsection{0735+178}
No emission lines have yet been identified in the optical spectrum of 0735+178,
though a lower limit to the redshift of 0.424 is provided by a pair of
absorption lines \citep{carswell74}. VLBI images of this source show a compact
core and a jet that extends to the northeast (e.g. \citealp{gabuzda92}) while
arcsecond scale VLA images reveal a point source
\citep{ulvestad83,kollgaard92}. 8, 22 and 43~GHz VLBI images reveal a jet out to
approximately 10~mas with two sharp apparent bends of 90$^{\circ}$ within 2~mas
of the core. It also appears that there was a dramatic change in the trajectory
of the jet within the first few mas between 1992 and 1995
\citep{gomez99,gomez01b}. This could be interpreted either as a precessing jet
nozzle or more likely as a result of pressure gradients in the external medium
through which the jet propagates \citep{gomez01b}.

The images presented here show, for the first time, coherent structures in this
source on scales of several tens of mas. The sharp twists close to the core
reported at higher frequencies are not seen here owing to the low resolution,
but the jet appears to turn from its initial easterly direction towards the
south at a distance of about 30~mas from the core. The beginning of this bend
is apparent in the 2.3~GHz, 1.7~GHz and 610~MHz images. In the 327~MHz image
there is a suggestion of emission 100~mas due south of the core which may
indicate that the jet continues on its southerly path after the initial turn at
30~mas from the core, although the detection is not strong enough to confirm
this.

The flux density in the previously undetected structures, beyond 10~mas from
the core, is 0.122 Jy, which, after accounting for the spectral index effects,
represents a very large fraction of the predicted intermediate scale flux at
5~GHz of $\sim 0.090$~Jy \citep{gabuzda94}.

At the lower limit for the redshift of this object, 1 mas $ \sim 4.9$ pc. The
luminosity detected in the new, extended, structure in this observation is
therefore $> 6.1 \times 10^{25}$ W Hz$^{-1}$.

\begin{figure}
\centering
\includegraphics[width=6.5cm]{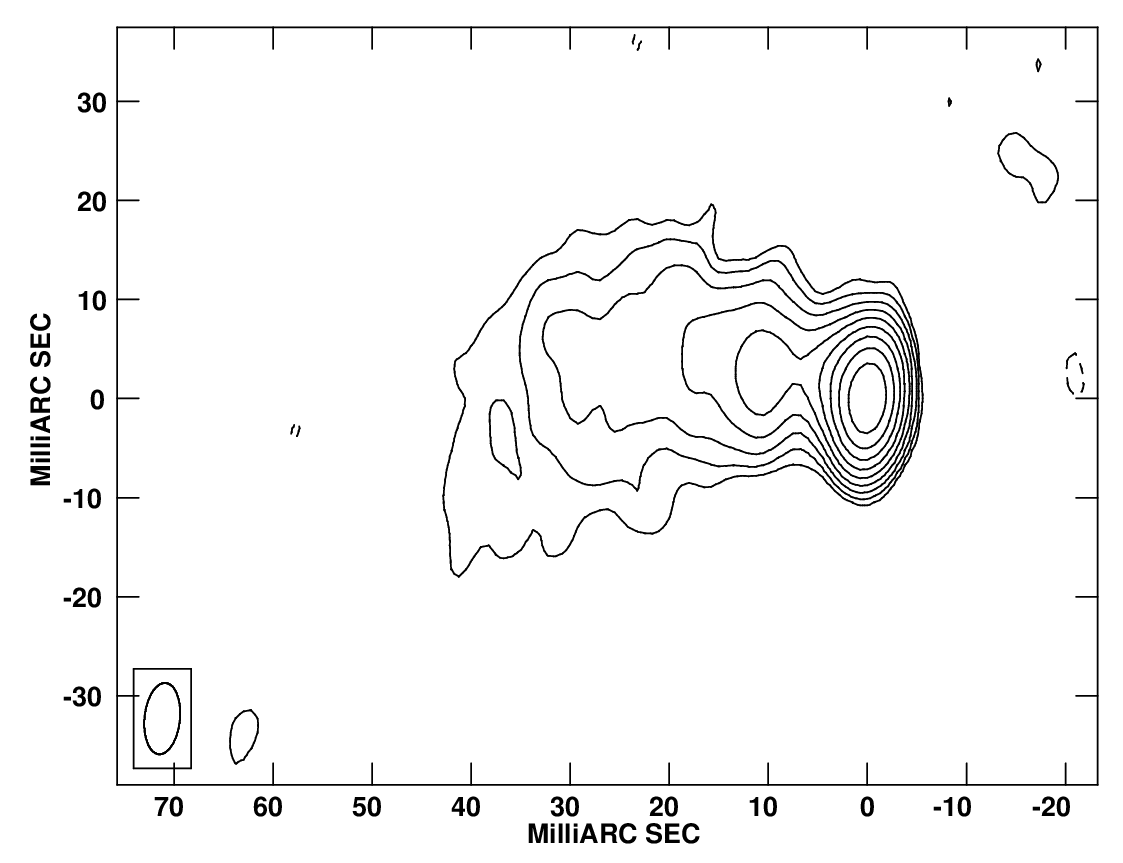}
   \caption{2.3~GHz image of 0735+178. The peak flux density is 0.78 Jy/beam
and the contour levels are $-0.2$, 0.2, 0.4, 0.8 ... 51.2 \%.
           }
      \label{fig:0735+178_2.3ghz}
\end{figure}

\begin{figure}
\centering
\includegraphics[width=6.5cm]{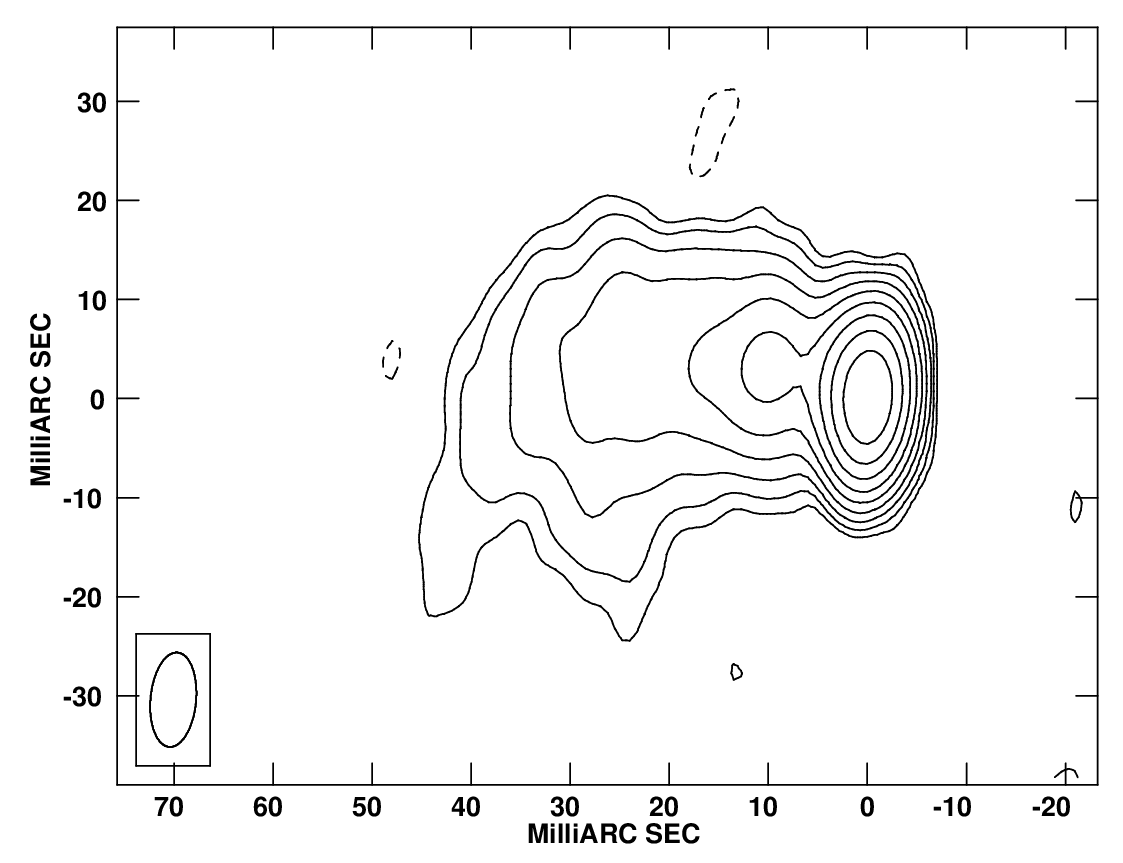}
   \caption{1.7~GHz image of 0735+178. The peak flux density is 0.79 Jy/beam
and the contour levels are $-0.2$, 0.2, 0.4, 0.8 ... 51.2 \%.
           }
      \label{fig:0735+178_1.7ghz}
\end{figure}

\begin{figure}
\centering
\includegraphics[width=6.5cm]{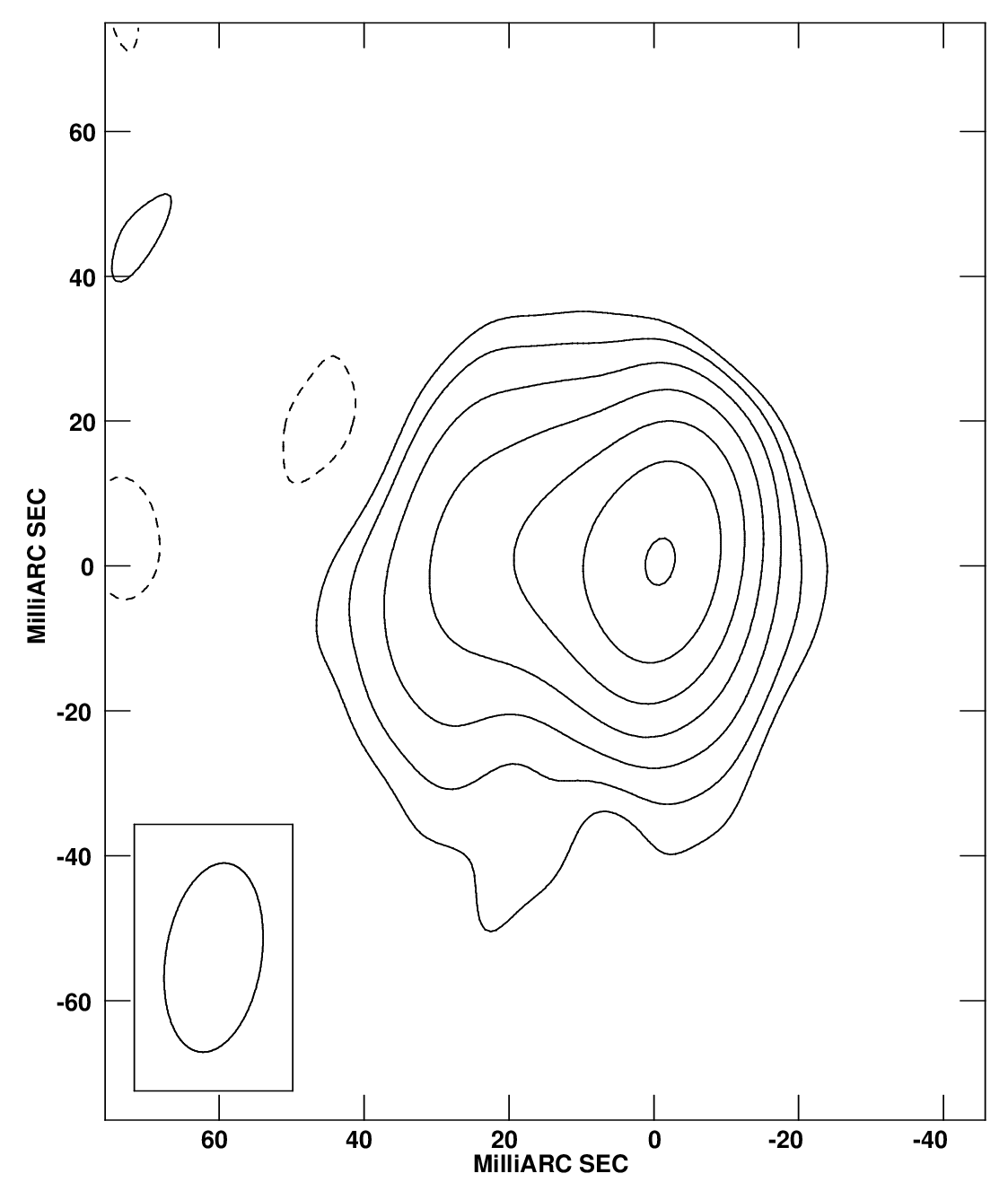}
   \caption{610~MHz image of 0735+178. The peak flux density is 0.65 Jy/beam
and the contour levels are $-1.5$, 1.5, 3.0, 6.0 ... 96.0 \%.
           }
      \label{fig:0735+178_0.6ghz}
\end{figure}

\begin{figure}
\centering
\includegraphics[width=6.5cm]{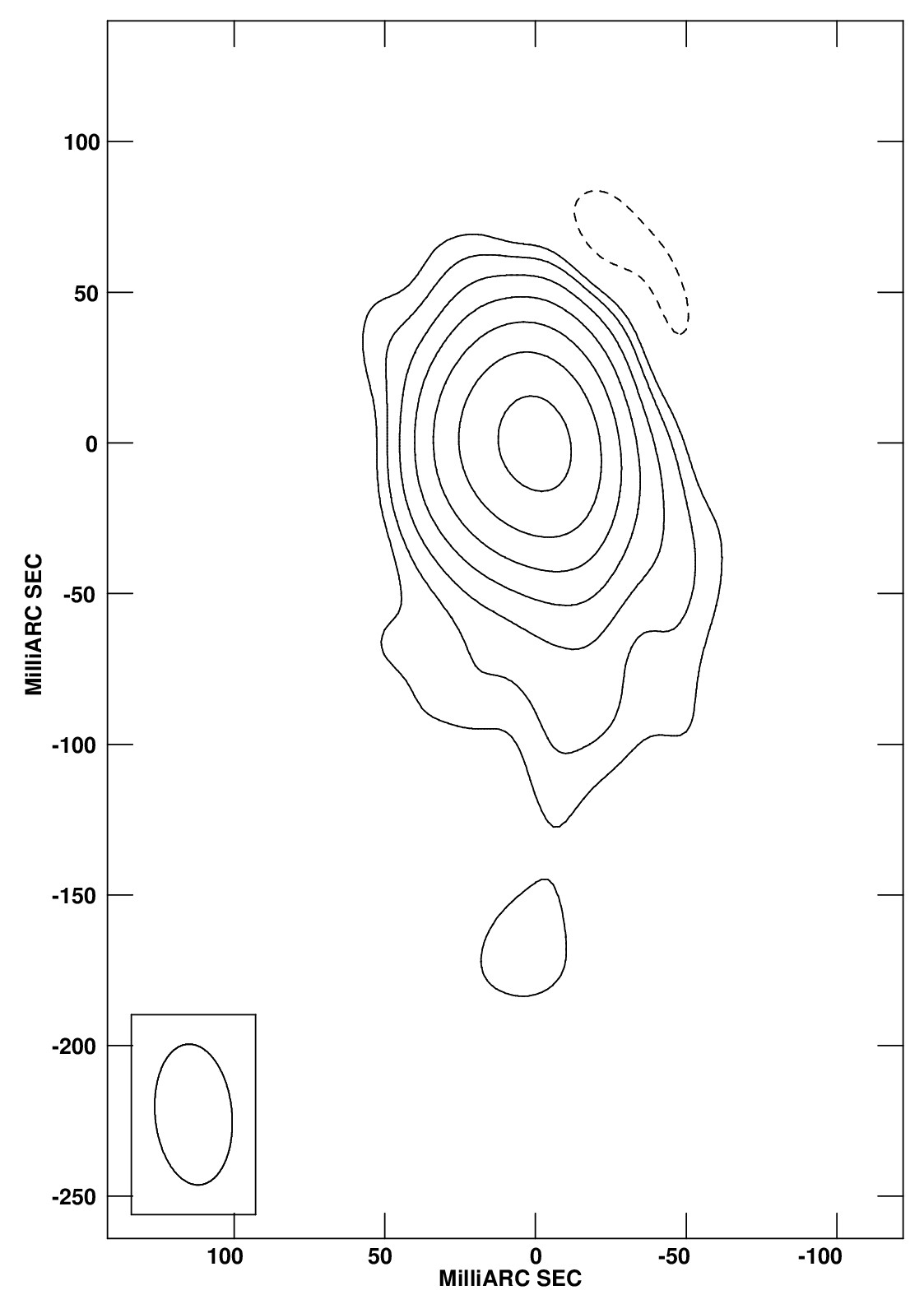}
   \caption{327~MHz image of 0735+178. The peak flux density is 0.43 Jy/beam
and the contour levels are $-1.2$, 1.2, 2.4, 4.8 ... 76.8 \%.
           }
      \label{fig:0735+178_0.3ghz}
\end{figure}


\section{Discussion}
It has been suggested that BL~Lac objects are low luminosity (FR\,I) radio
galaxies orientated such that Doppler boosting enhances their apparent
brightness (e.g. \citealt{browne83}). If this model is valid, the extended
structure in BL~Lacs should have a similar radio power to that found in the
extended structure of FR\,I radio galaxies. The FR\,I/FR\,II luminosity
division is $\sim 10^{24.5}$ -- $10^{25}$ W Hz$^{-1}$ at 1.4 GHz
\citep{bridle84b}. Allowing for a spectral index of $\sim -0.7$, this would
correspond to a 2.3~GHz division at approximately $\sim 2.0 \times 10^{24}$ --
$6.3 \times 10^{24}$ W Hz$^{-1}$.

For the two objects presented here the detected intermediate scale flux density
was a large fraction of that expected from the discrepancy between previous
5~GHz VLA and VLBI images. In the case of 0735+178, this may well account for
almost all of the intermediate scale ``missing" flux. The intermediate scale
flux detected for 1219+285 is somewhat less than the predicted intermediate
scale flux which means that more flux may lie on larger scales than are probed
in this observation. 

The luminosity of the intermediate scale emission detected in 1219+285 is $\sim
2.2 \times 10^{24}$ W Hz$^{-1}$. This emission appears to be rather diffuse,
and might be associated with a plume-like structure formed by the disruption of
the well-collimated parsec-scale jet seen closer to the core. Even if all of
this emission is associated with a diffuse unbeamed component, it is at the
lower end of the transition region between FR\,I and FR\,II radio sources, and
so does not pose a problem for associating the diffuse emission in this galaxy
with FR\,I structure. 

The intermediate scale flux density detected in 0735+178 appears to be
associated with a well-collimated jet, and is therefore likely to be
significantly Doppler beamed. For this reason, it is probably not significant
that the observed intermediate scale luminosity is above the FR\,I limit.

\section{Conclusions}
\begin{enumerate}
\item
In the two sources presented here, coherent structures extend out to
projected distances of $> 30$~pc from the core before disruption.
\item
In one case, 1219+285, diffuse -- possibly plume-like -- emission was
detected.
\item
The luminosity detected in the intermediate scale structure of 1219+285
presented here was found to be at the lower end of the
FR\,I/FR\,II luminosity division. 0735+178 had a luminosity in the
extended structure above this division, but as the emission appears to be
associated with a jet rather than a diffuse component this does not necessarily
imply that the source contains FR\,II extended structure.
\end{enumerate}

\begin{acknowledgements}
The National Radio Astronomy Observatory is operated by Associated Universities
Incorporated, under cooperative agreement with the NSF.
\end{acknowledgements}


\end{document}